\documentclass[pra,twocolumn,showpacs,amsmath,amssymb,superscriptaddress,eqsecnum]{revtex4-1}
\usepackage{graphicx,bm,color,mathptmx,hyperref} 
\newcommand{\Tr}{\mathop{\mathrm{Tr}} \nolimits}

\newcommand{\opx}{\hat{\sigma}_x}
\newcommand{\opy}{\hat{\sigma}_y}
\newcommand{\opz}{\hat{\sigma}_z}
\newcommand{\identity}{\hat{\openone}}

\begin{document}

\title{Sizing up entanglement in mutually unbiased bases with Fisher information}

\author{J.~\v{R}eh\'{a}\v{c}ek}
\affiliation{Department of Optics, Palack\'y University,
17. listopadu 12, 771 46 Olomouc, Czech Republic}

\author{Z.~Hradil}
\affiliation{Department of Optics, Palack\'y University,
17. listopadu 12, 771 46 Olomouc, Czech Republic}

\author{A.~B.~Klimov}
\affiliation{Departamento de F\'{\i}sica, Universidad de Guadalajara,
 44420~Guadalajara, Jalisco, Mexico}

\author{G.~Leuchs}
\affiliation{Max-Planck-Institut f\"ur die Physik des Lichts, 
G\"{u}nther-Scharowsky-Stra{\ss}e 1, Bau 24, 
91058 Erlangen, Germany}
\affiliation{Department f\"{u}r Physik, 
Universit\"{a}t Erlangen-N\"{u}rnberg,
Staudtstra{\ss}e 7, Bau 2, 91058 Erlangen, Germany}

\author{L.~L.~S\'{a}nchez-Soto} 
\affiliation{Max-Planck-Institut f\"ur die Physik des Lichts, 
G\"{u}nther-Scharowsky-Stra{\ss}e 1, Bau 24, 
91058 Erlangen, Germany}
\affiliation{Department f\"{u}r Physik, 
Universit\"{a}t Erlangen-N\"{u}rnberg,
Staudtstra{\ss}e 7, Bau 2, 91058 Erlangen, Germany}
\affiliation{Departamento de \'Optica, Facultad de F\'{\i}sica,
Universidad Complutense, 28040~Madrid, Spain}

\begin{abstract}
  An efficient method for assessing the quality of quantum state
  tomography is developed. Special attention is paid to the tomography
  of multipartite systems in terms of unbiased measurements.  Although
  the overall reconstruction errors of different sets of mutually
  unbiased bases are the same, differences appear when particular
  aspects of the measured system are contemplated.  This point is
  illustrated by estimating the fidelities of genuinely tripartite
  entangled states. 
\end{abstract}

\pacs{03.65.Wj, 03.65.Ta, 03.65.Aa, 03.67.Mn}

\maketitle

\section{Introduction}

Uncertainty in quantum theory can be attributed to two different
issues: the irreducible indeterminacy of individual quantum processes,
postulated by Born~\cite{Born:1926zr}, and the complementarity,
introduced by Bohr~\cite{Bohr:1935ly}, which implies that we cannot
simultaneously perform precise measurements of noncommuting
observables. For that reason, estimation of the quantum state from the
measurement outcomes, which is sometimes called quantum tomography, is
of paramount importance~\cite{Helstrom:1976ij,Holevo:2003fv,
  lnp:2004uq}.  Moreover, in practice, unavoidable imperfections and
finite resources come into play and the performance of tomographic
schemes should be assessed and compared.

At a fundamental level, mutually unbiased bases (MUBs) provide perhaps
the most accurate statement of complementarity. This idea emerged in
the seminal work of
Schwinger~\cite{Schwinger:1960a,*Schwinger:1960b,*Schwinger:1960c} and
it has gradually turned into a keystone of quantum information. Apart
from being instrumental in many hard problems~\cite{Durt:2010hc}, MUBs
have long been known to provide an optimal basis for quantum
tomography~\cite{Wootters:1989vn}.

When the Hilbert space dimension $d$ is a prime power, $d = p^{n}$, it
is known that there exist sets of $d+1$ MUBs~\cite{Ivanovic:1981tg}.
These Hilbert spaces are realized by systems of  $n$ particles,
which, in turn, allow for special entanglement properties.  More
specifically, we consider here $n$ qubits, as they are the building
blocks of quantum information processing. It was known
that for three qubits, four different set of MUBs exist with very
different entanglement properties~\cite{Lawrence:2002ij}. The analysis
was further extended to higher number of qubits~\cite{Romero:2005dz}
and confirmed by different
approaches~\cite{Lawrence:2011fu,Wiesniak:2011kl}.  For the
experimentalist, this information is very important, because the
complexity of an implementation of a given set of MUBs will, of
course, greatly depend on how many of the qubits need to be entangled.
   
In this work, we use Fisher information to set forth efficient tools
for assessing the quality of a wide class of tomographic schemes,
paying special attention to $n$-qubit MUBs.  Despite the widespread
belief that MUBs are equally sensitive to all the state features, we
show that MUBs with different entanglement properties differ in their
potential to characterize particular aspects of the reconstructed
state. We illustrate this with the fidelity estimation of three-qubit
states, making evident the relevance of MUB-entanglement
classification and illustrating the possibility to optimize MUB
tomography with respect to the observables of interest.

\section{Setting the scenario}

Let us start with a brief outlook of the basic methods we need for the
rest of the discussion. We deal with a $d$-dimensional quantum system,
represented by a positive semidefinite $d \times d$ density matrix
$\varrho$.  A very convenient parametrization of $\varrho$ can be
achieved in terms of a traceless Hermitian operator basis
$\{\lambda_{k}\}$, satisfying $\Tr (\lambda_{k} )= 0$ and $\Tr
(\lambda_{k} \lambda_{\ell}) = \delta_{k \ell}$:
\begin{equation}
  \label{rhodecomp}
  \varrho (\mathbf{a} ) =\frac{\openone}{d} + 
  \sum_{k=1}^{d^2-1} a_{k} \, \lambda_{k} \, ,
\end{equation}
where the $(d^{2}-1)$-dimensional generalized Bloch vector $\mathbf{a}
= ( a_{1}, \ldots, a_{d^{2}-1})$ ($a_{k} \in \mathbb{R}$) is uniquely
determined by $a_{k} = \Tr (\lambda_{k} \varrho)$.  The set $\{
\lambda_{k} \}$ coincides with the orthogonal generators of SU($d$),
which is the associated symmetry
algebra~\cite{Hioe:1981ve,Kimura:2003qf}.

The measurements performed on the system are described, in general, by
positive operator-valued measures (POVMs), which are a collection of
positive operators $\{ \Pi_{j} \ge 0 \}$ , resolving the identity
$\sum_{j} \Pi_{j}= \openone$~\cite{Holevo:2003fv}.  Each POVM element
represents a single output channel of the measuring apparatus; the
probability of detecting the $j$th output is given by the Born rule
$p_{j} = \Tr ( \varrho \Pi_{j})$.

In a sensible estimation procedure, we have $N$ identical copies of
the system and repeat the measurement on each of them. The statistics
of the outcomes is then multinomial, i.e.,
\begin{equation}
  \label{eq:multi}
  P (\mathbf{n}|\mathbf{a})  \propto \prod p_{j}^{n_{j}} \, ,
\end{equation} 
where $p_{j}$ is the probability of detection at the $j$th channel and
$n_{j}$ the actual number of detections. Here, $ P
(\mathbf{n}|\mathbf{a}) $ is the probability of registering the data
$\mathbf{n}$ provided the true state is $\mathbf{a}$.

The estimation requires the introduction of an
estimator; i.e., a rule of inference that allows one to extract a
value for $\mathbf{a}$ from the outcomes $\mathbf{n}$.  The random
variable $\hat{\mathbf{a}}$ is an unbiased estimator if 􏰋$\langle
\hat{\mathbf{a}} \rangle = \mathbf{a}$.  The ultimate bound on the
precision with which one can estimate $\mathbf{a}$ is given by the
Cramer-Rao bound~\cite{Cramer:1946ye,Rao:1973qo}, which, in terms of
the covariance matrix $\mathrm{cov}_{k \ell} (\hat{a} ) = \langle
\hat{a}_{k} \hat{a}_{\ell} \rangle - \langle \hat{a}_{k} \rangle
\langle \hat{a}_{\ell} \rangle$, can be stated as
\begin{equation}
  \label{CRbound}
  \mathrm{cov}_{k\ell} ( \hat{\mathbf{a}} ) \ge  
  (\mathrm{F}^{-1}) _{k\ell} \, ,
\end{equation} 
where F stands for the Fisher matrix~\cite{Fisher:1922tw}
\begin{equation}
  \label{fish-indep}
  \mathrm{F}_{k\ell}={N} \sum_{j} \frac{1}{p_{j}}
  \frac{\partial p_{j}}{\partial a_{k}}
  \frac{\partial p_{j}}{\partial a_{\ell}} \, . 
\end{equation}
Since $\partial p_{j}/\partial a_{k}=\Tr ( \lambda_{k} \Pi_{j} ) $, it
might superficially appear that computing the Fisher matrix (and hence
the covariances of the estimated parameters) is
straightforward. However, in practice, this can become quite a
difficult task: the cost of computing $M\times M$ matrix
multiplications and inversions, even with the best-known exact
algorithm~\cite{Strassen:1969bs}, scales as $O(M^{2.8})$. For a system
of $n$ qubits, this computational cost goes as $O(2^{5.6\,n})$, which
sets an upper limit on the dimension for which the evaluation of the
reconstruction errors is feasible. For instance, analyzing a system of
just five qubits requires about a billion of arithmetic operations with
individual elements, which makes the problem intractable along these
lines.  This especially applies when a large number of repeated
evaluations is required, as in Monte Carlo simulations.

This numerical cost can be considerably reduced by employing a special
parametrization for $\varrho$. To this end, we restrict ourselves to
informationally complete (IC) measurements, which are those for which
the outcome probabilities are sufficient to determine an arbitrary
quantum state~\cite{Prugovecki:1977fk,Busch:1989kx,
  Ariano:2004kx,Sych:2012zt}.  Given a system of dimension $d$, any IC
measurement must have at least $d^2$ output channels; when it has just
$m = d^2$ of them, it will be called a minimal IC reconstruction
scheme. It turns out that the error analysis in this case is
particularly simple and can be done analytically avoiding
time-expensive computations. We remark that we are not addressing here
the resources needed for a complete tomography (which scale
exponentially); rather, our aim is to ascertain what can be better
estimated from IC measurements.

Indeed, for an IC minimal scheme $\{ \Pi_{j} \}$ ($j=1, \ldots, m$),
there exists a unique representation of any quantum state in terms of
the basic probabilities $p_{j} = \Tr (\varrho \Pi_{j})$:
\begin{equation}
  \varrho = \varrho (p_{1}, p_{2}, \ldots, p_{m}) \, . 
\end{equation}
Normalization reduces by one the number of independent parameters
describing the state: $a_{k} \equiv p_{k} $ ($k=1, \ldots, m-1)$ and
$p_{m} = 1-\sum_{k=1}^{m-1} a_{k}$. In this way, we get $\partial
p_{j} / \partial a_{k} =\delta_{j k}$ ($j=1, \ldots, m-1$) and
$\partial p_{m}/\partial a_{k}=-1$, leading us to
\begin{equation}
\label{fisher inverse}
  \textrm{F}_{k \ell}=\frac{1}{ p_{k}}   \delta_{k\ell} +   
 \frac{1}{p_{m}} , 
  \quad 
 (\textrm{F}^{-1})_{k \ell}  =  p_{k}  (1-p_{k} )  \delta_{k \ell}  - 
 p_{k} p_{\ell} (1- \delta_{k\ell}) \, .
\end{equation}

We thus conclude that the errors of any IC minimal scheme
are given by the covariance matrix of the underlying true multinomial
distribution governing the measurement outcomes.  Notice that this
might not apply to other overdetermined setups, as, for instance, 
optical homodyne tomography.

Once the Fisher matrix is known, the errors in any observation can be
estimated. Let us consider the measurement of the average value $z=\Tr
(\varrho Z)$ of some generic observable $Z$. If the reconstructed
state is $\hat{\rho}$, the predicted outcomes are $\hat{z}=\Tr (
\hat{\varrho} Z)$ and the expected errors are
\begin{equation}
  \label{error}
  (\Delta z)^2= \left \langle 
    \Tr [ Z ( \varrho - \hat{\varrho} ) ]^2 \right \rangle \, ,
\end{equation}
where the averaging here is over many repetitions of the
reconstruction.  By expanding the observable in the POVM elements,
$Z=\sum_{k}^m z_k \Pi_k$, the true and predicted outcomes can be given
in terms of true and predicted measurement probabilities as
$z=\sum_k^m z_k p_k$ and $\hat z=\sum_k^m z_k \hat{p}_k$, where
$p_k=\Tr(\rho \Pi_k)$ and $\hat{p}_k=\Tr(\hat\rho \Pi_k)$,
respectively.  Denoting $\Delta p_k=p_k-\hat{p}_k$ the differences
between the true and inferred probabilities, the expected error of the
observable $Z$ defined by Eq.~\eqref{error} becomes
\begin{equation}
  \label{error_p}
  (\Delta z)^2=\sum_{k\ell}^m z_k \langle \Delta p_k \Delta p_{\ell}
  \rangle z_\ell \, .
\end{equation}
Next, we rearrange Eq.~\eqref{error_p} so that only $m-1$ linearly
independent probabilities $p_1, \ldots, p_{m-1}$ are involved. Notice
that $\sum_k^m p_k=\sum_k^m \hat p_k=1$ implies $\sum_k^m \Delta
p_k=0$ and hence $\Delta p_m= -\sum_k^{m-1} \Delta p_k$. In this way,
we get
\begin{eqnarray}
  (\Delta z)^2 & = &  z_m^2 \langle (\Delta p_m)^2\rangle + \sum_{k
    \ell}^{m-1} 
  z_k z_\ell \langle \Delta p_k \Delta p_\ell \rangle \nonumber \\
  &+ &  \sum_k^{m-1} z_k z_m \langle \Delta p_k \Delta p_m\rangle +
  \sum_\ell^{m-1} z_m z_\ell \langle \Delta p_m \Delta p_\ell\rangle
  \nonumber \\
  & =& \sum_{k\ell}^{m-1} (z_k -z_m)  \langle \Delta p_k \Delta p_\ell\rangle (z_\ell-z_m)
\end{eqnarray}
Employing the Cramer-Rao lower bound, we finally obtain
\begin{equation}
  \label{error_final}
  (\Delta z)^2=\sum_{k\ell}^{m-1}(z_k -z_m) ( \mathrm{F}^{-1})_{k \ell} (z_\ell-z_m).
\end{equation}
where $\mathrm{F}^{-1}$ is the inverse Fisher matrix in the
probability representation given by Eq.~\eqref{fisher inverse}.

Occasionally, working in the measurement representation might be
preferable.  Expanding the true and estimated states in the measured
POVM elements, $\varrho=\sum_{k}^{m} w_{k} \Pi_{k}$ and $\hat{\varrho}
= \sum_{k}^{m} \hat{w}_{k} \Pi_{k}$, we seek to express $(\Delta z)^2$
in terms of the reconstruction errors $\Delta w_k= w_{k}-\hat{w}_{k}$.
If
\begin{equation}
  J_{k\ell}=\frac{\partial p_{k}}{\partial w_{\ell}}, \qquad
  k,\ell=1,\ldots , m-1 
\end{equation}
is the matrix connecting the measurement and probability
representations, then $\Delta p_k=\sum_\ell^{m-1} J_{k\ell} \; \Delta
w_\ell$ and we have $\mathrm{F}^{-1} = J^T \mathrm{F}_w^{-1} J$, which
is computed effectively provided  $J$ is sparse.  By
inserting this into Eq.~\eqref{error_final}, we get the desired
result.

\section{Assessing MUB performance}

One pertinent example for which this probability representation
turns out to be very efficient is for MUBs. As heralded
before, we consider a system of $n$ qubits; since the dimension
$d=2^n$ is a power of a prime sets of MUBs $\{ | \Psi_{\alpha, j}
\rangle \}$ exist and explicit construction procedures are at
hand~\cite{Durt:2010hc}.  We denote the corresponding projectors by
$\Pi_{\alpha ,j}$ where the Greek index $\alpha$ labels one of the $d
+ 1$ families of MUBs and $j$ denotes one of the $d$ orthogonal states
in this family. Unbiasedness translates into
\begin{equation}
  \Tr  (  \Pi_{\alpha j  }   \Pi_{\beta k}  )  =  
  \delta_{\alpha \beta}  \delta_{j k }  +  
  \frac{1}{d} (1 -  \delta_{\alpha \beta})  \, .
\end{equation}
Notice that, in agreement with the usual MUB terminology, each set of
eigenstates is normalized to unity $\sum_j^d\Pi_{\alpha j}=1$, so that
the POVM becomes normalized to $d+1$, rather than to unity. Our
previous convention is readily recovered by using $N(d+1)$ as the
total number of copies.

As the total number of projections [$d(d+1)$] minus the number of
constraints [$(d+1)$] matches the minimal number [$(d^2-1)$] of
independent measurements, this MUB tomography is indeed an IC minimal
scheme.  In addition, there are $d+1$ sets of vectors, each resolving
the unity, so that $\sum_{j} p_{\alpha j}=1$ holds for each observable
$\alpha$.  Consequently, the Fisher information matrix in the
probability representation takes a block diagonal form
\begin{equation}
  \mathrm{F} = \bigoplus_\alpha^{d+1} \mathrm{F}_{\alpha} \, ,
\end{equation} 
and in each block $\alpha$, $(\mathrm{F}_{\alpha})_{k \ell}$ and
$(\mathrm{F}_{\alpha}^{-1})_{k \ell}$ are given by Eq.~\eqref{fisher
  inverse}.

Measurement errors can now be easily estimated. Indeed, expanding a
generic state in the MUB basis as
\begin{equation}
  \varrho =  \sum_\alpha^{d+1} \sum_j^d w_{\alpha  j } \Pi_{\alpha j}
  \, ,
\end{equation}
with $ w_{\alpha j }= p_{\alpha j } - 1/(d+1)$, we find that
$J_{\alpha k,\beta \ell}=\delta_{\alpha \beta}\delta_{k\ell}$ and the
total error appears as a sum of independent contributions $(\Delta
z)^2 = \sum_\alpha (\Delta z_\alpha)^2$ of individual MUB eigensets,
with
\begin{equation}
\label{eq:uf}
  (\Delta z_\alpha)^2=\sum_{k \ell}^{d-1} (z_{\alpha k}-z_{\alpha m}) 
  \big( \mathrm{F}_{\alpha}\big)^{-1}_{k \ell} 
  (z_{\alpha \ell}-z_{\alpha m}) \, .
\end{equation}

We can reinterpret these results in an alternative way. Without lack
of generality, we consider one diagonal block and drop the index
$\alpha$. We introduce the operator $S$ by $S |z\rangle = |\tilde{z}
\rangle$, with $z= (z_{1}, \ldots, z_{d} )$, $ \tilde{z}= (
\tilde{z}_{1}, \ldots, \tilde{z}_{d-1} )$, and $\tilde{z}_{k}=z_{k} -
z_{d}$.  This can be represented by the rectangular $(d-1)\times
d$-dimensional matrix $S_{kj}=\delta_{k j}-\delta_{j d}$.  If $
\langle \cdot | \cdot \rangle$ is the standard scalar product in this
$d$-dimensional vector space, we may simply write
\begin{equation}
  \label{scalarproduct}
  (\Delta z)^2 = \langle z| \tilde{\mathrm{F}}^{-1}| z\rangle, 
\end{equation}
where $\tilde{\mathrm{F}}^{-1}=S^T \mathrm{F}^{-1}S$.  Notice that the
inferred variance takes now the form of Born's rule and the effective
inverse Fisher matrix $\tilde{\mathrm{F}}^{-1}$ becomes the relevant
object governing the errors.  For example, the mean Hilbert-Schmidt
distance from the estimate to the true state becomes
\begin{equation}
  \langle \Tr [  ( \rho - \hat \rho)^2] \rangle=
  \Tr (\tilde{\mathrm{F}}^{-1}) \, ,
\end{equation}
which gives a unitary invariant error, as it might be
anticipated~\cite{Zhu:2011bh}.

Another instance of interest is the error of the fidelity measurement
$Z=| \Psi_{\mathrm{true}} \rangle \langle \Psi_{\mathrm{true}} |
$. Here
\begin{equation}
  z_{\alpha k} = w_{\alpha k}= p_{\alpha k} - \frac{1}{d+1}\, .
\end{equation}
The constant term $1/(d+1)$ can be dropped because $z_{\alpha k}$
enters Eq.~(\ref{eq:uf}) only through differences.  These formulas,
together with Eq.~\eqref{fisher inverse}, provide timely tools for
analyzing the performance and optimality of different MUB
reconstruction schemes.

For example, one could be interested in which observable can be most
(least) accurately inferred from a MUB tomography. This is tantamount
to minimizing (maximizing) Eq.~\eqref{scalarproduct} subject to a
fixed norm of $|z\rangle$.  The principal axes of the error ellipsoid
are defined by the eigenvectors and eigenvalues of the effective
inverse Fisher matrix $\tilde{\mathrm{F}}^{-1}$.  Notice that there is
always one zero eigenvalue per diagonal block, corresponding to a
constant vector $z_{k}=\text{const}$ ($ k=1,\ldots,d)$, which, in
turn, corresponds to measuring the trace of the signal density matrix
$Z\propto 1$ and $z=\Tr (\rho)$. This is consistent with the fact that
the trace is constrained to unity and hence error free. Thus, the
least (most) favorable measurements from the point of view of a
particular detection scheme are those given by the largest (second
smallest) eigenvectors of $\tilde{\mathrm{F}}^{-1}$.

Similarly, one may be interested in the distribution $P[(\Delta z)^2]$
of errors among all possible inferred measurements.  From
Eq.~(\ref{scalarproduct}), such a distribution is given precisely by
the restricted numerical range or ``real shadow''~\cite{Dunkl:2011dq}
of $\tilde{\mathrm{F}}^{-1}$.

\section{The case of three-qubits}

Once the formalism has been set up, let us see how it works for MUB
reconstruction schemes. The goal is to assess the performance of
different sets of MUBs for moderately-sized quantum systems.

For definiteness, let us look at the case of three qubits.  It is
known~\cite{Romero:2005dz} that MUB sets can be divided into
nonequivalent classes with respect to entanglement properties. In the
eight-dimensional Hilbert space of three qubits, any complete set is
comprised of 9 MUBs. We label the different sets by $(n_1,n_2,n_3)$,
where $n_{1}$ denotes the number of separable bases (every eigenvector
of theses bases is a tensor product of singe-qubit states), $n_{2}$
the number of biseparable bases (one qubit is factorized and the other
two are in a maximally entangled state) and $n_{3}$ the number of
nonseparable bases. In this notation, there are four classes:
$(0,9,0)$, $(3,0,6)$, $(2,3,4)$, and $(1,6,2)$ (see the supplemental
material for a detailed description).  On physical grounds, one could
expect that the performances of these four classes with respect to
entanglement-specific state properties will also be different.

To clarify the question we apply the procedure developed so far. To
tie the observable to entanglement properties, we consider the
estimation of the fidelity error taking the inferred observable to be
the projection on the true state, $Z = |\Psi_{\mathrm{true}} \rangle
\langle \Psi_{\mathrm{true}}|$. The confidence in the inferred
fidelity can be used at the same time as a simple criterion of the
overall quality of the tomographic setup.

\begin{figure}
  \includegraphics[width=0.95\columnwidth]{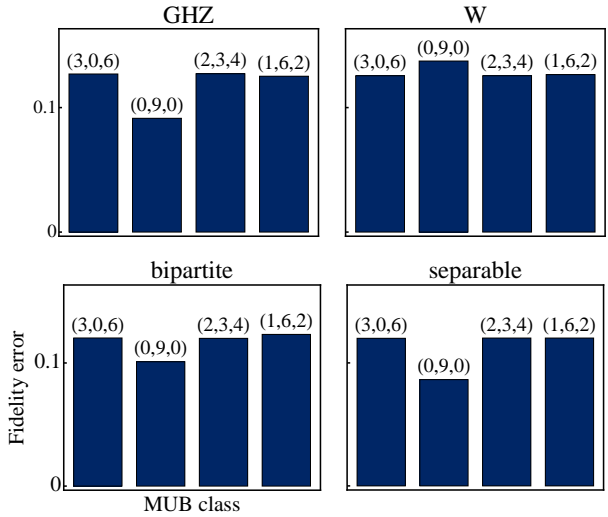}
  \caption{Fidelity errors of estimated GHZ and W states (upper panel)
    as well  as bipartite and separable states (lower panel) for MUB
    tomography measurements. Notice  that the bipartite-bases
    $(0,9,0)$ class performs best on GHZ, bipartite and separable
    states,  but worst on W states. The differences  are up to $30\%$.
    \label{fig1}}
\end{figure}

In our simulations, genuinely tripartite GHZ and W entangled
three-qubit states were randomly generated, as well as bipartite and
separable ones, and subjected to MUB measurements.  For each MUB
class, the expected error of the inferred fidelity was estimated and
averaged over $20,000$ randomly distributed states. The result is
shown in Fig.~\ref{fig1}. As we can appreciate, in average the
$(0,9,0)$ scheme yields smaller fidelity errors for GHZ-like states
(for bipartite and separable also) and larger errors for W states
compared to other MUB sets. One also notices that the difference
between GHZ, bipartite and separable is small, while W are still
different.

\begin{figure}
  \includegraphics[width=0.95\columnwidth]{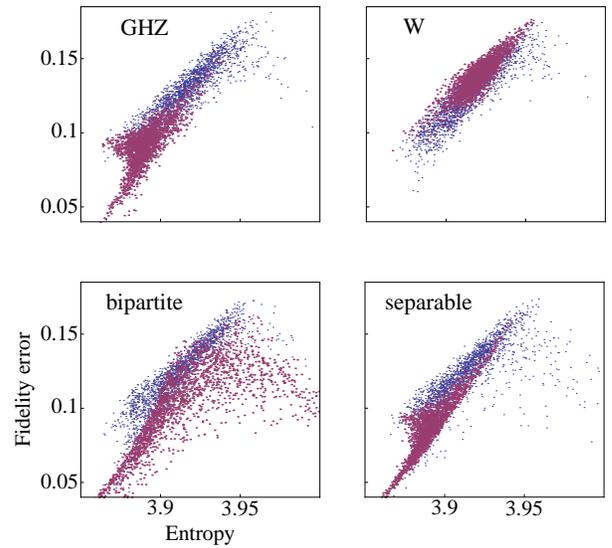}
  \caption{Correlations between the entropy of tomography outcomes and
    the corresponding  fidelity errors for the same states as in Fig.~1. Large red and smalle blue
    dots are for $(0,9,0)$ and $(3,0,6)$ bases,  respectively.  
    \label{fig2}}
\end{figure}

To understand this result better, we have also investigated the noise
in the measured data and calculated the correlations between the
entropy of measurement probabilities and fidelity errors, restricting
ourselves, for simplicity, to GHZ and W states. Figure~\ref{fig2}
confirms strong correlations between those two magnitudes and shows
that for GHZ states, $(0,9,0)$ tomograms are less noisy on average
than $(3,0,6)$ tomograms. Again, the opposite is true for W states.
There is not a simple connection between
state entanglement and MUB classes, as one might naively
anticipate. This demonstrates the key role played by error estimation
along the lines presented here.

\begin{figure}[b]
  \includegraphics[width=\columnwidth]{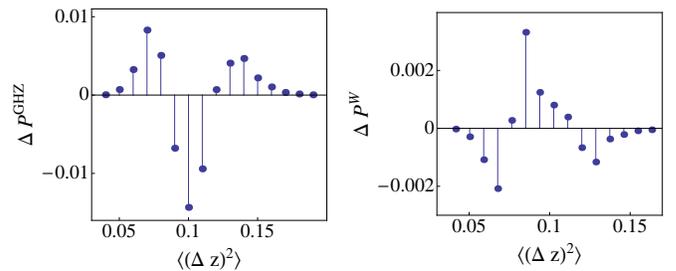}
\caption{Differences of $(0,9,0)$ and $(3,0,6)$ histograms of
  variances  for GHZ and W states.
All histograms are normalized to unity. Positive $\Delta P$ for some
$(\Delta z)^2$ means error of this  size is more typical for $(0,9,0)$
scheme than for $(3,0,6)$ scheme and  \textit{vice versa}. 
\label{fig3}}
\end{figure}

Finally, we consider a set of randomly chosen measurements $Z_{i}$,
whose expectation values $\hat{z}_{i} = \Tr ( \hat{\rho} Z_i)$ are
inferred from the estimated state $\hat{\rho}$ and analyze the
probability distribution $P[(\Delta z)^2]$ of the expected variances $
\langle (\Delta z_i)^2\rangle$. This distribution, being the real
shadow $\langle z_i|\tilde{\mathrm{F}}^{-1}|z_i\rangle$ of the
effective inverse Fisher matrix, tells us how the errors are
distributed in the set of all possible measurements and hence
describes in detail the performance of a given reconstruction scheme.
In Fig.~\ref{fig3}, the difference $\Delta P=P_{(0,9,0)}[(\Delta
z)^2]-P_{(3,0,6)}[(\Delta z)^2]$ of the $(0,9,0)$ and $(3,0,6)$
shadows is approximated by histograms obtained from $500,000$
variances calculated for randomly generated measurements $|z_i\rangle$
and randomly choosen GHZ and W states for each MUB set.  For GHZ
states, the $(0,9,0)$ scheme yields very small errors and very large
errors more often than the $(3,0,6)$ scheme. The opposite is true for
W states.  Such complementary behavior of different MUB sets makes it
possible to optimize the tomography setup.  For instance, quantities
with low inherent noise are more precisely determined by the $(0,9,0)$
[$(3,0,6)$] set provided the true state is GHZ-like (W-like). The
fidelity measurement discussed above is a salient example of that
optimization.

We stress that this kind of analysis, where many repeated evaluations
of the Fisher matrix must be performed, is chiefly suited to the
proposed method, for the standard approach would become quickly
unfeasible due to the scaling with the system dimension.

\section{Concluding remarks}

To summarize, we have presented a complete Fisher-information-based
toolbox for the proper assessment of any IC tomographic scheme. In
this context, we have discussed the case of set of MUBs; although all
of them are IC, their performance for some particular tasks turns out
to be dependent on the entanglement distribution among the states of
the set. We believe that the ideas and techniques developed here will
be relevant in the experimental implementation and optimization of
quantum protocols in higher-dimensional Hilbert spaces.

\begin{acknowledgments}
Financial support from the EU FP7 (Grant Q-ESSENCE), the Spanish DGI
(Grant FIS2011-26786), the UCM-BSCH program (Grant GR-920992), the
Mexican CONACyT (Grant 106525), the Czech Ministry of Trade and
Industry (Grant FR-TI1/384) and the Technology Agency of the Czech
Republic (Grant TE01020229) is acknowledged.
\end{acknowledgments}


\begin{thebibliography}{27}%
\makeatletter
\providecommand \@ifxundefined [1]{%
 \@ifx{#1\undefined}
}%
\providecommand \@ifnum [1]{%
 \ifnum #1\expandafter \@firstoftwo
 \else \expandafter \@secondoftwo
 \fi
}%
\providecommand \@ifx [1]{%
 \ifx #1\expandafter \@firstoftwo
 \else \expandafter \@secondoftwo
 \fi
}%
\providecommand \natexlab [1]{#1}%
\providecommand \enquote  [1]{``#1''}%
\providecommand \bibnamefont  [1]{#1}%
\providecommand \bibfnamefont [1]{#1}%
\providecommand \citenamefont [1]{#1}%
\providecommand \href@noop [0]{\@secondoftwo}%
\providecommand \href [0]{\begingroup \@sanitize@url \@href}%
\providecommand \@href[1]{\@@startlink{#1}\@@href}%
\providecommand \@@href[1]{\endgroup#1\@@endlink}%
\providecommand \@sanitize@url [0]{\catcode `\\12\catcode `\$12\catcode
  `\&12\catcode `\#12\catcode `\^12\catcode `\_12\catcode `\%12\relax}%
\providecommand \@@startlink[1]{}%
\providecommand \@@endlink[0]{}%
\providecommand \url  [0]{\begingroup\@sanitize@url \@url }%
\providecommand \@url [1]{\endgroup\@href {#1}{\urlprefix }}%
\providecommand \urlprefix  [0]{URL }%
\providecommand \Eprint [0]{\href }%
\providecommand \doibase [0]{http://dx.doi.org/}%
\providecommand \selectlanguage [0]{\@gobble}%
\providecommand \bibinfo  [0]{\@secondoftwo}%
\providecommand \bibfield  [0]{\@secondoftwo}%
\providecommand \translation [1]{[#1]}%
\providecommand \BibitemOpen [0]{}%
\providecommand \bibitemStop [0]{}%
\providecommand \bibitemNoStop [0]{.\EOS\space}%
\providecommand \EOS [0]{\spacefactor3000\relax}%
\providecommand \BibitemShut  [1]{\csname bibitem#1\endcsname}%
\let\auto@bib@innerbib\@empty
\bibitem [{\citenamefont {Born}(1926)}]{Born:1926zr}%
  \BibitemOpen
  \bibfield  {author} {\bibinfo {author} {\bibfnamefont {M.}~\bibnamefont
  {Born}},\ }\href@noop {} {\bibfield  {journal} {\bibinfo  {journal} {Z.
  Phys.}\ }\textbf {\bibinfo {volume} {37}},\ \bibinfo {pages} {863} (\bibinfo
  {year} {1926})}\BibitemShut {NoStop}%
\bibitem [{\citenamefont {Bohr}(1935)}]{Bohr:1935ly}%
  \BibitemOpen
  \bibfield  {author} {\bibinfo {author} {\bibfnamefont {N.}~\bibnamefont
  {Bohr}},\ }\href@noop {} {\bibfield  {journal} {\bibinfo  {journal} {Phys.
  Rev.}\ }\textbf {\bibinfo {volume} {48}},\ \bibinfo {pages} {696} (\bibinfo
  {year} {1935})}\BibitemShut {NoStop}%
\bibitem [{\citenamefont {Helstrom}(1976)}]{Helstrom:1976ij}%
  \BibitemOpen
  \bibfield  {author} {\bibinfo {author} {\bibfnamefont {C.~W.}\ \bibnamefont
  {Helstrom}},\ }\href@noop {} {\emph {\bibinfo {title} {Quantum Detection and
  Estimation Theory}}}\ (\bibinfo  {publisher} {Academic},\ \bibinfo {address}
  {New York},\ \bibinfo {year} {1976})\BibitemShut {NoStop}%
\bibitem [{\citenamefont {Holevo}(2003)}]{Holevo:2003fv}%
  \BibitemOpen
  \bibfield  {author} {\bibinfo {author} {\bibfnamefont {A.~S.}\ \bibnamefont
  {Holevo}},\ }\href@noop {} {\emph {\bibinfo {title} {Probabilistic and
  Statistical Aspects of Quantum Theory}}},\ \bibinfo {edition} {2nd}\ ed.\
  (\bibinfo  {publisher} {North Holland},\ \bibinfo {address} {Amsterdam},\
  \bibinfo {year} {2003})\BibitemShut {NoStop}%
\bibitem [{\citenamefont {Paris}\ and\ \citenamefont
  {\v{R}eh\'a\v{c}ek}(2004)}]{lnp:2004uq}%
  \BibitemOpen
  \bibinfo {editor} {\bibfnamefont {M.~G.~A.}\ \bibnamefont {Paris}}\ and\
  \bibinfo {editor} {\bibfnamefont {J.}~\bibnamefont {\v{R}eh\'a\v{c}ek}},\
  eds.,\ \href@noop {} {\emph {\bibinfo {title} {Quantum State Estimation}}},\
  \bibinfo {series} {Lect. Not. Phys.}, Vol.\ \bibinfo {volume} {649}\
  (\bibinfo  {publisher} {Springer},\ \bibinfo {address} {Berlin},\ \bibinfo
  {year} {2004})\BibitemShut {NoStop}%
\bibitem [{\citenamefont {Schwinger}(1960{\natexlab{a}})}]{Schwinger:1960a}%
  \BibitemOpen
  \bibfield  {author} {\bibinfo {author} {\bibfnamefont {J.}~\bibnamefont
  {Schwinger}},\ }\href@noop {} {\bibfield  {journal} {\bibinfo  {journal}
  {Proc. Natl. Acad. Sci. USA}\ }\textbf {\bibinfo {volume} {46}},\ \bibinfo
  {pages} {570} (\bibinfo {year} {1960}{\natexlab{a}})}\BibitemShut {NoStop}%
\bibitem [{\citenamefont {Schwinger}(1960{\natexlab{b}})}]{Schwinger:1960b}%
  \BibitemOpen
  \bibfield  {author} {\bibinfo {author} {\bibfnamefont {J.}~\bibnamefont
  {Schwinger}},\ }\href@noop {} {\bibfield  {journal} {\bibinfo  {journal}
  {Proc. Natl. Acad. Sci. USA}\ }\textbf {\bibinfo {volume} {46}},\ \bibinfo
  {pages} {883} (\bibinfo {year} {1960}{\natexlab{b}})}\BibitemShut {NoStop}%
\bibitem [{\citenamefont {Schwinger}(1960{\natexlab{c}})}]{Schwinger:1960c}%
  \BibitemOpen
  \bibfield  {author} {\bibinfo {author} {\bibfnamefont {J.}~\bibnamefont
  {Schwinger}},\ }\href@noop {} {\bibfield  {journal} {\bibinfo  {journal}
  {Proc. Natl. Acad. Sci. USA}\ }\textbf {\bibinfo {volume} {46}},\ \bibinfo
  {pages} {1401} (\bibinfo {year} {1960}{\natexlab{c}})}\BibitemShut {NoStop}%
\bibitem [{\citenamefont {Durt}\ \emph {et~al.}(2010)\citenamefont {Durt},
  \citenamefont {Englert}, \citenamefont {Bengtsson},\ and\ \citenamefont
  {Zyczkowski}}]{Durt:2010hc}%
  \BibitemOpen
  \bibfield  {author} {\bibinfo {author} {\bibfnamefont {T.}~\bibnamefont
  {Durt}}, \bibinfo {author} {\bibfnamefont {B.-G.}\ \bibnamefont {Englert}},
  \bibinfo {author} {\bibfnamefont {I.}~\bibnamefont {Bengtsson}}, \ and\
  \bibinfo {author} {\bibfnamefont {K.}~\bibnamefont {Zyczkowski}},\
  }\href@noop {} {\bibfield  {journal} {\bibinfo  {journal} {Int. J. Quantum
  Inf.}\ }\textbf {\bibinfo {volume} {8}},\ \bibinfo {pages} {535} (\bibinfo
  {year} {2010})}\BibitemShut {NoStop}%
\bibitem [{\citenamefont {Wootters}\ and\ \citenamefont
  {Fields}(1989)}]{Wootters:1989vn}%
  \BibitemOpen
  \bibfield  {author} {\bibinfo {author} {\bibfnamefont {W.~K.}\ \bibnamefont
  {Wootters}}\ and\ \bibinfo {author} {\bibfnamefont {B.~D.}\ \bibnamefont
  {Fields}},\ }\href@noop {} {\bibfield  {journal} {\bibinfo  {journal} {Ann.
  Phys.}\ }\textbf {\bibinfo {volume} {191}},\ \bibinfo {pages} {363} (\bibinfo
  {year} {1989})}\BibitemShut {NoStop}%
\bibitem [{\citenamefont {Ivanovic}(1981)}]{Ivanovic:1981tg}%
  \BibitemOpen
  \bibfield  {author} {\bibinfo {author} {\bibfnamefont {I.~D.}\ \bibnamefont
  {Ivanovic}},\ }\href@noop {} {\bibfield  {journal} {\bibinfo  {journal} {J.
  Phys. A}\ }\textbf {\bibinfo {volume} {14}},\ \bibinfo {pages} {3241}
  (\bibinfo {year} {1981})}\BibitemShut {NoStop}%
\bibitem [{\citenamefont {Lawrence}\ \emph {et~al.}(2002)\citenamefont
  {Lawrence}, \citenamefont {\v{C}. Brukner},\ and\ \citenamefont
  {Zeilinger}}]{Lawrence:2002ij}%
  \BibitemOpen
  \bibfield  {author} {\bibinfo {author} {\bibfnamefont {J.}~\bibnamefont
  {Lawrence}}, \bibinfo {author} {\bibnamefont {\v{C}. Brukner}}, \ and\
  \bibinfo {author} {\bibfnamefont {A.}~\bibnamefont {Zeilinger}},\ }\href@noop
  {} {\bibfield  {journal} {\bibinfo  {journal} {Phys. Rev. A}\ }\textbf
  {\bibinfo {volume} {65}},\ \bibinfo {pages} {032320} (\bibinfo {year}
  {2002})}\BibitemShut {NoStop}%
\bibitem [{\citenamefont {Romero}\ \emph {et~al.}(2005)\citenamefont {Romero},
  \citenamefont {Bj{\"{o}}rk}, \citenamefont {Klimov},\ and\ \citenamefont
  {S{{\'a}}nchez-Soto}}]{Romero:2005dz}%
  \BibitemOpen
  \bibfield  {author} {\bibinfo {author} {\bibfnamefont {J.~L.}\ \bibnamefont
  {Romero}}, \bibinfo {author} {\bibfnamefont {G.}~\bibnamefont {Bj{\"{o}}rk}},
  \bibinfo {author} {\bibfnamefont {A.~B.}\ \bibnamefont {Klimov}}, \ and\
  \bibinfo {author} {\bibfnamefont {L.~L.}\ \bibnamefont
  {S{{\'a}}nchez-Soto}},\ }\href@noop {} {\bibfield  {journal} {\bibinfo
  {journal} {Phys. Rev. A}\ }\textbf {\bibinfo {volume} {72}},\ \bibinfo
  {pages} {062310} (\bibinfo {year} {2005})}\BibitemShut {NoStop}%
\bibitem [{\citenamefont {Lawrence}(2011)}]{Lawrence:2011fu}%
  \BibitemOpen
  \bibfield  {author} {\bibinfo {author} {\bibfnamefont {J.}~\bibnamefont
  {Lawrence}},\ }\href@noop {} {\bibfield  {journal} {\bibinfo  {journal}
  {Phys. Rev. A}\ }\textbf {\bibinfo {volume} {84}},\ \bibinfo {pages} {022338}
  (\bibinfo {year} {2011})}\BibitemShut {NoStop}%
\bibitem [{\citenamefont {Wie{\'s}niak}\ \emph {et~al.}(2011)\citenamefont
  {Wie{\'s}niak}, \citenamefont {Paterek},\ and\ \citenamefont
  {Zeilinger}}]{Wiesniak:2011kl}%
  \BibitemOpen
  \bibfield  {author} {\bibinfo {author} {\bibfnamefont {M.}~\bibnamefont
  {Wie{\'s}niak}}, \bibinfo {author} {\bibfnamefont {T.}~\bibnamefont
  {Paterek}}, \ and\ \bibinfo {author} {\bibfnamefont {A.}~\bibnamefont
  {Zeilinger}},\ }\href@noop {} {\bibfield  {journal} {\bibinfo  {journal} {New
  J. Phys.}\ }\textbf {\bibinfo {volume} {13}},\ \bibinfo {pages} {053047}
  (\bibinfo {year} {2011})}\BibitemShut {NoStop}%
\bibitem [{\citenamefont {Hioe}\ and\ \citenamefont
  {Eberly}(1981)}]{Hioe:1981ve}%
  \BibitemOpen
  \bibfield  {author} {\bibinfo {author} {\bibfnamefont {F.~T.}\ \bibnamefont
  {Hioe}}\ and\ \bibinfo {author} {\bibfnamefont {J.~H.}\ \bibnamefont
  {Eberly}},\ }\href@noop {} {\bibfield  {journal} {\bibinfo  {journal} {Phys.
  Rev. Lett.}\ }\textbf {\bibinfo {volume} {47}},\ \bibinfo {pages} {838}
  (\bibinfo {year} {1981})}\BibitemShut {NoStop}%
\bibitem [{\citenamefont {Kimura}(2003)}]{Kimura:2003qf}%
  \BibitemOpen
  \bibfield  {author} {\bibinfo {author} {\bibfnamefont {G.}~\bibnamefont
  {Kimura}},\ }\href@noop {} {\bibfield  {journal} {\bibinfo  {journal} {Phys.
  Lett. A}\ }\textbf {\bibinfo {volume} {314}},\ \bibinfo {pages} {339}
  (\bibinfo {year} {2003})}\BibitemShut {NoStop}%
\bibitem [{\citenamefont {Cram{\'e}r}(1946)}]{Cramer:1946ye}%
  \BibitemOpen
  \bibfield  {author} {\bibinfo {author} {\bibfnamefont {H.}~\bibnamefont
  {Cram{\'e}r}},\ }\href@noop {} {\emph {\bibinfo {title} {Mathematical Methods
  of Statistics}}}\ (\bibinfo  {publisher} {Princeton University},\ \bibinfo
  {address} {Princeton},\ \bibinfo {year} {1946})\BibitemShut {NoStop}%
\bibitem [{\citenamefont {Rao}(1973)}]{Rao:1973qo}%
  \BibitemOpen
  \bibfield  {author} {\bibinfo {author} {\bibfnamefont {C.~R.}\ \bibnamefont
  {Rao}},\ }\href@noop {} {\emph {\bibinfo {title} {Linear Statistical
  Inference and Its Applications}}}\ (\bibinfo  {publisher} {Wiley},\ \bibinfo
  {address} {New York},\ \bibinfo {year} {1973})\BibitemShut {NoStop}%
\bibitem [{\citenamefont {Fisher}(1922)}]{Fisher:1922tw}%
  \BibitemOpen
  \bibfield  {author} {\bibinfo {author} {\bibfnamefont {R.~A.}\ \bibnamefont
  {Fisher}},\ }\href@noop {} {\bibfield  {journal} {\bibinfo  {journal} {Phil.
  Trans. R. Soc. A}\ }\textbf {\bibinfo {volume} {222}},\ \bibinfo {pages}
  {309} (\bibinfo {year} {1922})}\BibitemShut {NoStop}%
\bibitem [{\citenamefont {Strassen}(1969)}]{Strassen:1969bs}%
  \BibitemOpen
  \bibfield  {author} {\bibinfo {author} {\bibfnamefont {V.}~\bibnamefont
  {Strassen}},\ }\href@noop {} {\bibfield  {journal} {\bibinfo  {journal}
  {Numer. Math.}\ }\textbf {\bibinfo {volume} {13}},\ \bibinfo {pages} {354}
  (\bibinfo {year} {1969})}\BibitemShut {NoStop}%
\bibitem [{\citenamefont {Prugove\v{c}ki}(1977)}]{Prugovecki:1977fk}%
  \BibitemOpen
  \bibfield  {author} {\bibinfo {author} {\bibfnamefont {E.}~\bibnamefont
  {Prugove\v{c}ki}},\ }\href@noop {} {\bibfield  {journal} {\bibinfo  {journal}
  {Int. J. Theor. Phys.}\ }\textbf {\bibinfo {volume} {16}},\ \bibinfo {pages}
  {321} (\bibinfo {year} {1977})}\BibitemShut {NoStop}%
\bibitem [{\citenamefont {Busch}\ and\ \citenamefont
  {Lahti}(1989)}]{Busch:1989kx}%
  \BibitemOpen
  \bibfield  {author} {\bibinfo {author} {\bibfnamefont {P.}~\bibnamefont
  {Busch}}\ and\ \bibinfo {author} {\bibfnamefont {P.~J.}\ \bibnamefont
  {Lahti}},\ }\href@noop {} {\bibfield  {journal} {\bibinfo  {journal} {Found.
  Phys.}\ }\textbf {\bibinfo {volume} {19}},\ \bibinfo {pages} {633} (\bibinfo
  {year} {1989})}\BibitemShut {NoStop}%
\bibitem [{\citenamefont {Ariano}\ \emph {et~al.}(2004)\citenamefont {Ariano},
  \citenamefont {Perinotti},\ and\ \citenamefont {Sacchi}}]{Ariano:2004kx}%
  \BibitemOpen
  \bibfield  {author} {\bibinfo {author} {\bibfnamefont {G.~M.~D.}\
  \bibnamefont {Ariano}}, \bibinfo {author} {\bibfnamefont {P.}~\bibnamefont
  {Perinotti}}, \ and\ \bibinfo {author} {\bibfnamefont {M.~F.}\ \bibnamefont
  {Sacchi}},\ }\href@noop {} {\bibfield  {journal} {\bibinfo  {journal} {J.
  Opt. B}\ }\textbf {\bibinfo {volume} {6}},\ \bibinfo {pages} {S487} (\bibinfo
  {year} {2004})}\BibitemShut {NoStop}%
\bibitem [{\citenamefont {Sych}\ \emph {et~al.}(2012)\citenamefont {Sych},
  \citenamefont {{\v R}eh{\'a}{\v c}ek}, \citenamefont {Hradil}, \citenamefont
  {Leuchs},\ and\ \citenamefont {S{\'a}nchez-Soto}}]{Sych:2012zt}%
  \BibitemOpen
  \bibfield  {author} {\bibinfo {author} {\bibfnamefont {D.}~\bibnamefont
  {Sych}}, \bibinfo {author} {\bibfnamefont {J.}~\bibnamefont {{\v R}eh{\'a}{\v
  c}ek}}, \bibinfo {author} {\bibfnamefont {Z.}~\bibnamefont {Hradil}},
  \bibinfo {author} {\bibfnamefont {G.}~\bibnamefont {Leuchs}}, \ and\ \bibinfo
  {author} {\bibfnamefont {L.~L.}\ \bibnamefont {S{\'a}nchez-Soto}},\
  }\href@noop {} {\bibfield  {journal} {\bibinfo  {journal} {Phys. Rev. A}\
  }\textbf {\bibinfo {volume} {86}},\ \bibinfo {pages} {052123} (\bibinfo
  {year} {2012})}\BibitemShut {NoStop}%
\bibitem [{\citenamefont {Zhu}\ and\ \citenamefont
  {Englert}(2011)}]{Zhu:2011bh}%
  \BibitemOpen
  \bibfield  {author} {\bibinfo {author} {\bibfnamefont {H.}~\bibnamefont
  {Zhu}}\ and\ \bibinfo {author} {\bibfnamefont {B.-G.}\ \bibnamefont
  {Englert}},\ }\href {http://link.aps.org/doi/10.1103/PhysRevA.84.022327}
  {\bibfield  {journal} {\bibinfo  {journal} {Phys. Rev. A}\ }\textbf {\bibinfo
  {volume} {84}},\ \bibinfo {pages} {022327} (\bibinfo {year}
  {2011})}\BibitemShut {NoStop}%
\bibitem [{\citenamefont {Dunkl}\ \emph {et~al.}(2011)\citenamefont {Dunkl},
  \citenamefont {Gawron}, \citenamefont {Holbrook}, \citenamefont {Miszczak},
  \citenamefont {Pucha{\l}a},\ and\ \citenamefont
  {{\.Z}yczkowski}}]{Dunkl:2011dq}%
  \BibitemOpen
  \bibfield  {author} {\bibinfo {author} {\bibfnamefont {C.~F.}\ \bibnamefont
  {Dunkl}}, \bibinfo {author} {\bibfnamefont {P.}~\bibnamefont {Gawron}},
  \bibinfo {author} {\bibfnamefont {J.~A.}\ \bibnamefont {Holbrook}}, \bibinfo
  {author} {\bibfnamefont {J.~A.}\ \bibnamefont {Miszczak}}, \bibinfo {author}
  {\bibfnamefont {Z.}~\bibnamefont {Pucha{\l}a}}, \ and\ \bibinfo {author}
  {\bibfnamefont {K.}~\bibnamefont {{\.Z}yczkowski}},\ }\href@noop {}
  {\bibfield  {journal} {\bibinfo  {journal} {J. Phys. A}\ }\textbf {\bibinfo
  {volume} {44}},\ \bibinfo {pages} {335301} (\bibinfo {year}
  {2011})}\BibitemShut {NoStop}%
\end{thebibliography}
%

\newpage

\begin{widetext}


\section*{Explicit form of the set of MUBs for three qubits}

For the sake of completeness, we briefly review the structure of the
sets of MUBs for three qubits, following essentially the approach in Ref.~\cite{Romero:2005dz}.

Because states belonging to the same basis are usually taken
to be orthonormal, to study the property of ``mutually
unbiasedness" it is possible to use either mutually unbiased bases
or the operators which have the basis states as eigenvectors. We
thus need $d^2-1$ operators to obtain the whole set of
states. In the case of power of  prime dimension, this
set can be  constructed as $d+1$ classes of $d-1$ commuting
operators.

In this way we get tables with nine rows of three mutually commuting (tensor products of) operators.
We have suppressed the tensor multiplication sign in all the tables.
By construction, the simultaneous
eigenstates of the operators in each row give a complete basis,
and each basis is mutually unbiased to each other. The number on
the left enumerates the bases, while the number on the right
denotes how many subsystems the bases can be factorized into.

As stated before, we label the different
sets of MUBs by $(n_1,n_2,n_3)$, where $n_{1}$ denotes the number of
separable bases (every eigenvector of theses bases is a tensor product
of singe-qubit states), $n_{2}$ the number of biseparable bases (one
qubit is factorized and the other two are in a maximally entangled
state) and $n_{3}$ the number of nonseparable bases. 
In this way, the allowed structures are  $\{ (2,3,4), (0,9,0),
(1,6,2), (3,0,6) \}$ and the corresponding tables are given in the
following.

\begin{table}
\caption{(2,3,4) MUB.}
\begin{tabular}{ccccccccc} \hline \hline
1 & $\opz \identity \identity$ & $\identity \identity \opz$ &
$\identity \opz \identity$ & $\opz \identity \opz$ & $\identity
\opz \opz$ & $\opz \opz \opz$ & $\opz \opz \identity$ & 3 \\
2 & $\opx \identity \identity$ & $\identity \opx \identity $ &
$\identity \identity \opx$ & $\opx \opx \identity$ & $\identity
\opx \opx$ & $\opx \opx \opx$ & $\opx \identity \opx$ & 3 \\
3 & $\opy \identity \identity$ & $\identity \opx \opz$ &
$\identity \opz \opx$ & $\opy \opx \opz$ & $\identity \opy \opy$ &
$\opy \opy \opy$ & $\opy \opz \opx$ & 2 \\
4 & $\opx \identity \opz$ & $\identity \opy \identity$ &
$\opz \identity \opy $ & $\opx \opy \opz$ & $\opz \opy \opy$ &
$\opy \opy \opx$ & $\opy \identity \opx$ & 2 \\
5 & $\opx \opz \identity$ & $\opz \opx \opz$ &
$\identity \opz \opy$ & $\opy \opy \opz$ &
$\opz \opy \opx$ & $\opy \opx \opx$ & $\opx \identity \opy$ & 1\\
6 & $\opy \identity \opz$ & $\identity \opy \opz$ &
$\opz \opz \opy$ & $\opy \opy \identity$ &
$\opz \opx \opx$ & $\opx \opx \opy$ &
$\opx \opz \opx$ & 1 \\
7 & $\opx \opz \opz$ & $\opz \opy \opz$ &
$\opz \opz \opx$ & $\opy \opx \identity$ &
$\identity \opx \opy$ & $\opx \opy \opx$ &
$\opy \identity \opy$ & 1\\
8 & $\opy \opz \opz$ & $\opz \opy \identity$ &
$\opz \identity \opx$ & $\opx \opx \opz$ &
$\identity \opy \opx$ & $\opy \opx \opy$ &
$\opx \opz \opy$ & 1 \\
9 & $\opy \opz \identity$ & $\opz \opx \identity $ &
$\identity \identity \opy$ & $\opx \opy \identity$ &
$\opz \opx \opy$ & $\opx \opy \opy$ & $\opy \opz \opy$ & 2 \\
\hline \hline
\end{tabular}
\label{table1}
\end{table}

\begin{table}
\caption{(0,9,0) MUB.}
\begin{tabular}{ccccccccc} \hline \hline
1 & $\opx \identity \identity$ & $\identity \opz \opy$ &
$\identity \opy \opz$ & $\opx \opz \opy$ &
$\identity \opx \opx$ & $\opx \opx \opx$ &
$\opx \opy \opz$ & 2 \\
2 & $\opy \identity \identity$ & $\identity \opx \opz $ &
$\identity \opz \opx$ & $\opy \opx \opz$ &
$\identity \opy \opy$ & $\opy \opy \opy$ &
$\opy \opz \opx$ & 2 \\
3 & $\opz \identity \identity$ & $\identity \opy \opx$ &
$\identity \opx \opy$ & $\opz \opy \opx$ &
$\identity \opz \opz$ & $\opz \opz \opz$ &
$\opz \opx \opy$ & 2\\
4 & $\opy \opz \opy$ & $\identity \opz \identity$ &
$\opx \identity \opz $ & $\opy \identity \opy$ &
$\opx \opz \opz$ & $\opz \identity \opx$ &
$\opz \opz \opx$ & 2 \\
5 & $\opy \opy \opz$ & $\opx \opy \opx$ &
$\identity \opy \identity$ & $\opz \identity \opy$ &
$\opx \identity \opx$ & $\opz \opy \opy$ &
$\opy \identity \opz$ & 2 \\
6 & $\opz \opz \opy$ & $\identity \identity \opy$ &
$\opx \opy \identity$ & $\opz \opz \identity$ &
$\opx \opy \opy$ & $\opy \opx \identity$ &
$\opy \opx \opy$ & 2 \\ 
7 & $\opy \opx \opx$ & $\opx \identity \opy$ &
$\opx \opx \opy$ & $\opz \opx \opz$ &
$\identity \opx \identity$ & $\opy \identity \opx$ &
$\opz \identity \opz$ & 2 \\
8 & $\opz \opx \opx$ & $\opx \opz \identity$ &
$\opx \opz \opx$ & $\opy \opy \opx$ &
$\identity \identity \opx$ & $\opz \opx \identity$ &
$\opy \opy \identity$ & 2 \\
9 & $\opz \opy \opz$ & $\opx \opx \opz $ &
$\identity \identity \opz$ & $\opy \opz \identity$ &
$\opx \opx \identity$ & $\opy \opz \opz$ & $\opz
\opy \identity$ & 2 \\ \hline \hline
\end{tabular}
\label{table3}
\end{table}

\begin{table}
\caption{(1,6,2) MUB.}
\begin{tabular}{ccccccccc} \hline \hline
1 & $\opx \opz \identity$ & $\identity \identity \opy$ & $\opz \opy
\identity$ & $\opx \opz \opy$ & $\opz \opy \opy$ & $\opy \opx \opy$
& $\opy \opx \identity$ & 2\\ 
 2 & $\opy \opz \identity$ &
$\opz \opx \identity $ & $\identity \identity \opx$ & $\opx \opy
\identity$ &
$\opz \opx \opx$ & $\opx \opy \opx$ & $\opy \opz \opx$ & 2\\
3 & $\opz \identity \identity$ & $\opz \opx \opy$ & $\opz
\opy \opx$ & $\identity \opx \opy$ &
$\identity \opz \opz$ & $\opz \opz \opz$ & $\identity \opy \opx$ & 2\\
4 & $\opy \opz \opy$ & $\identity \opz \identity$ & $\opx
\opz \opz $ & $\opy \identity \opy$ &
$\opx \identity \opz$ & $\opz \opz \opx$ & $\opz \identity \opx$ & 2\\
5 & $\opx \opx \identity$ & $\opy \opy \opy$ & $\opz \opy
\opz$ & $\opz \opz \opy$ &
$\opx \identity \opx$ & $\identity \opx \opx$ & $\opy \opz \opz$ & 1\\
6 & $\opz \identity \opy$ & $\identity \opz \opy$ & $\opy
\opx \opz$ & $\opz \opz \identity$ & $\opy \opy \opx$ & $\opx \opy
\opz$ & $\opx \opx \opx$ & 1\\ 
7 & $\opx \opx \opy$ & $\opx
\identity \opy$ & $\opy \opx \opx$ & $\identity \opx
\identity$ & $\opz \opx \opz$ & $\opy \identity\opx$ & $\opz \identity \opz$ & 2\\
8 & $\identity \opy \opy$ & $\opx \identity \identity$ &
$\opx \opz \opx$ & $\opx \opy \opy$ & $\identity \opz
\opx$ & $\identity \opx \opz$ & $\opx \opx \opz$ & 2\\
9 & $\identity \opy \identity$ & $\opy \opy \identity $ &
$\identity \identity \opz$ & $\opy \identity \identity$ & $\opy \opy
\opz$ & $\opy \identity \opz$ & $\identity \opy \opz$ & 3\\
\hline \hline
\end{tabular}
\label{table4}
\end{table}

\begin{table}
\caption{(3,0,6) MUB.}
\begin{tabular}{ccccccccc} \hline \hline
1 & $\opx \opy \opz$ & $\identity \opz \opx$ & $\opy \opz \identity$
& $\opx \opx \opy$ & $\opy \identity \opx$ & $\opz \opy \opy$ &
$\opz \opx \opz$ & 1\\ 
2 & $\opz \opy \opz$ & $\opy \opx \opz
$ & $\identity \opz \opy$ & $\opx \opz \identity$ &
$\opy \opy \opx$ & $\opx \identity \opy$ & $\opz \opx \opx$ & 1\\
3 & $\opy \identity \identity$ & $\opy \opy \opy$ & $\opy
\identity \opy$ & $\identity \opy \opy$ &
$\identity \opy \identity$ & $\opy \opy \identity$ & $\identity \identity \opy$ & 3\\
4 & $\opz \opx \opy$ & $\identity \opy \opz$ & $\opx \opy
\identity $ & $\opz \opz \opx$ &
$\opx \identity \opz$ & $\opy \opx \opx$ & $\opy \opz \opy$ & 1\\
5 & $\opx \opx \opz$ & $\opz \identity \opx$ & $\opy \opz
\opz$ & $\opy \opx \opy$ &
$\opx \opz \opy$ & $\identity \opy \opx$ & $\opz \opy \identity$ & 1\\
6 & $\opy \opz \opx$ & $\identity \opx \opy$ & $\opz \opx
\identity$ & $\opy \opy \opz$ & $\opz \identity \opy$ & $\opx \opz
\opz$ & $\opx \opy \opx$ & 1\\  
7 & $\opx \opy \opy$ & $\opx
\opz \opx$ & $\opz \opy \opx$ & $\identity \opx
\opz$ & $\opy \opx \identity$ & $\opz \opz \opy$ & $\opy \identity \opz$ & 1\\
8 & $\identity \identity \opx$ & $\opx \identity \identity$ &
$\opx \opx \opx$ & $\opx \identity \opx$ & $\identity \opx
\opx$ & $\identity \opx \identity$ & $\opx \opx \identity$ & 3\\
9 & $\identity \opz \identity$ & $\opz \opz \identity $ &
$\identity \identity \opz$ & $\opz \identity \identity$ & $\opz \opz
\opz$ & $\opz \identity \opz$ & $\identity \opz \opz$ & 3\\
\hline \hline
\end{tabular}
\label{table6}
\end{table}

\end{widetext}

\end{document}